 \documentclass[twocolumn]{webofc}
\usepackage[varg]{txfonts}   
\usepackage{hyperref}
\usepackage{url}
\hypersetup{colorlinks=true,citecolor=blue,urlcolor=blue,linkcolor=blue}
\begin{document}
\title{Triplet pairing in neutron matter in a comprehensive diagrammatic approach}
%
%

\author{\firstname{Panagiota} \lastname{Papakonstantinou}\inst{1}\fnsep\thanks{\email{ppapakon@ibs.re.kr}} \and
        \firstname{Eckhard} \lastname{Krotscheck}\inst{2,3}\fnsep\thanks{\email{eckhardk@buffalo.edu}} \and
        \firstname{Jiawei} \lastname{Wang}\inst{2}\fnsep\thanks{\email{jwang97@buffalo.edu}}
}

\institute{Institute for Rare Isotope Science (IRIS), Institute for Basic Science(IBS), Daejeon 34000, Korea 
\and
          Department of Physics, SUNY Buffalo, New York 14260, USA 
\and
           Institute for Theoretical Physics, Johannes Kepler University, A 4040 Linz, Austria
          }

\abstract{We apply a large-scale summation of Feynman diagrams,
  including the class of parquet diagrams plus important contributions
  outside the parquet class, for calculating effective pairing
  interactions and subsequently the superfluid gap in $P-$wave pairing
  in neutron matter.  We use realistic nucleon-nucleon interactions of
  the $v_8$ type and perform calculations up to a Fermi momentum of
  $1.8$~fm$^{-1}$.  We find that many-body correlations lead to a
  strong reduction of the spin-orbit interaction, and, therefore, to a
  radical suppression of the $^3P_2-^3\!F_2$ gap and an enhancement of
  the $^3P_0$ gap.  }
\maketitle
\section{Introduction}
\label{intro}

The superfluid phase transition in neutron matter has been studied
microscopically for decades; see, for example,
Refs.~\cite{SeC2019,Str2018} for related reviews and compilations of
the relevant literature.  Numerous microscopic calculations predict
singlet-state superfluidity consistently at sub-saturation densities.
Triplet pairing is generally predicted at higher densities.  However,
predictions for the magnitude of the triplet-pairing gap vary widely.
Whether triplet pairing develops in dense matter is particular important in astrophysics, because it can affect the cooling of neutron stars in intricate ways, enhancing certain mechanisms and suppressing others. 

In this work, we undertake the most comprehensive to-date diagrammatic approach to calculating triplet pairing in dense neutron matter 
by including not only ring and ladder diagrams 
but also diagrams beyond the parquet class, which the form of the nuclear interaction necessitates. 
 
Next, we summarize existing microscopic approaches to the nuclear pairing gap, we briefly describe the formalism used in the present work, and we proceed to present our results.

\section{Existing approaches}
\label{existing}
We may classify existing microscopic approaches to calculating the pairing gap in nuclear matter as follows: 
First, there are calculations of the superfluid gap at the mean-field level
using bare, more or less realistic nuclear interactions. 
The effect of a three-body force has also been explored. 
Second, there are those that take into account medium polarization effects, which can strongly influence the superfluid transition temperature. 
Such calculations  necessarily involve assumptions about the quasiparticle interaction based on known physical observables. 
Third, inclusion of many-body effects in correlated basis function theory for superfluid systems, 
which can be mapped onto a
regular Bardeen–Cooper–Schrieffer (BCS)-like theory with
effective interactions. 
Finally, some Monte Carlo calculations exist for pairing in the singlet channel
in low-density neutron matter. 
The present work falls into the third category. 
We proceed to sketch our method before we present our results. 

\section{Correlated wavefunctions, diagrammatic summations, and the nuclear interaction} 
\label{correlated} 

\begin{figure} 
{\mbox{~}\hspace{7mm} (a)       \hspace{2.6cm}       (b)}\\ [-5mm] 
\centerline{
\includegraphics[width=2.75cm]{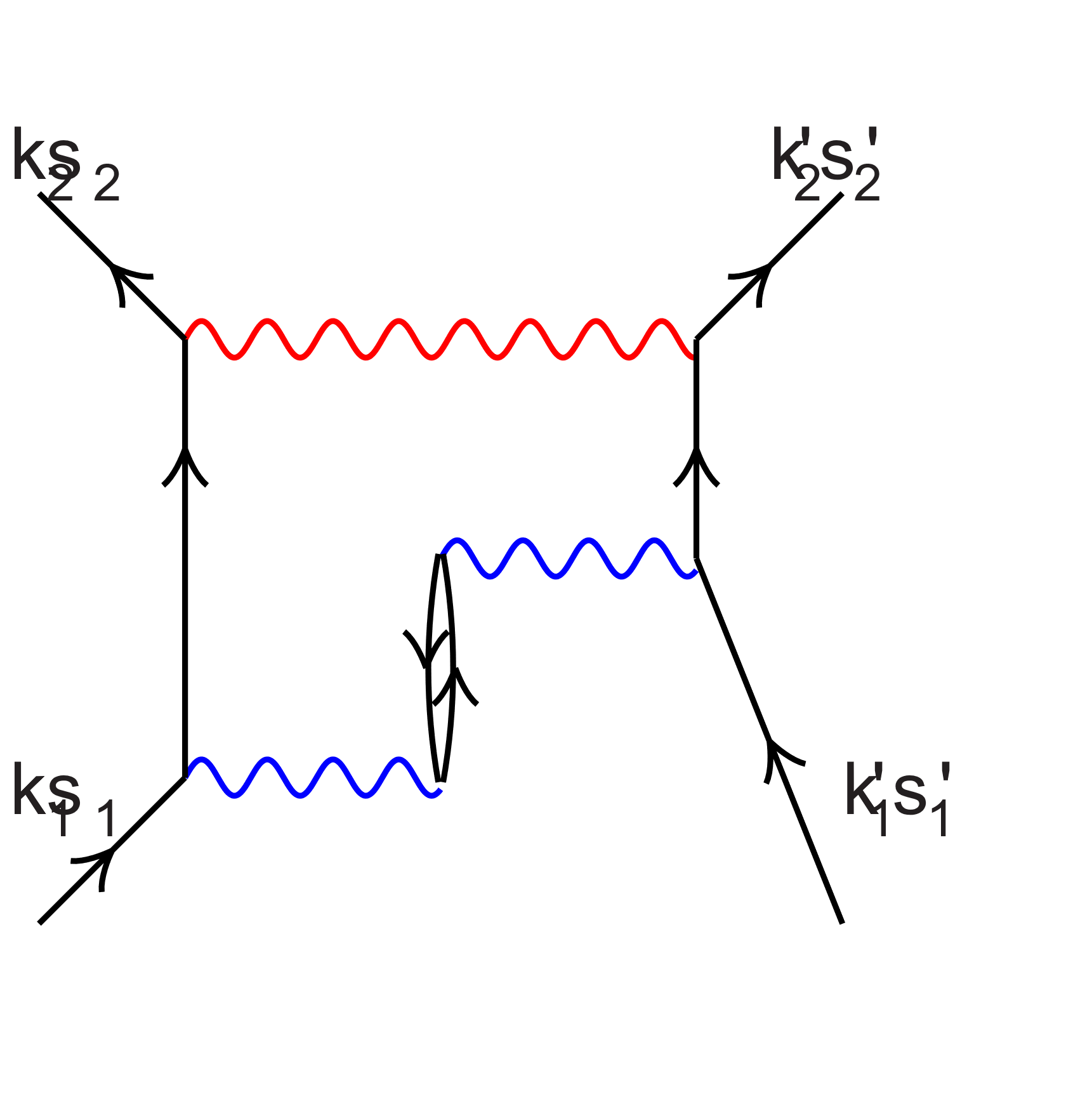} \hspace{3mm} 
\includegraphics[width=3.08cm]{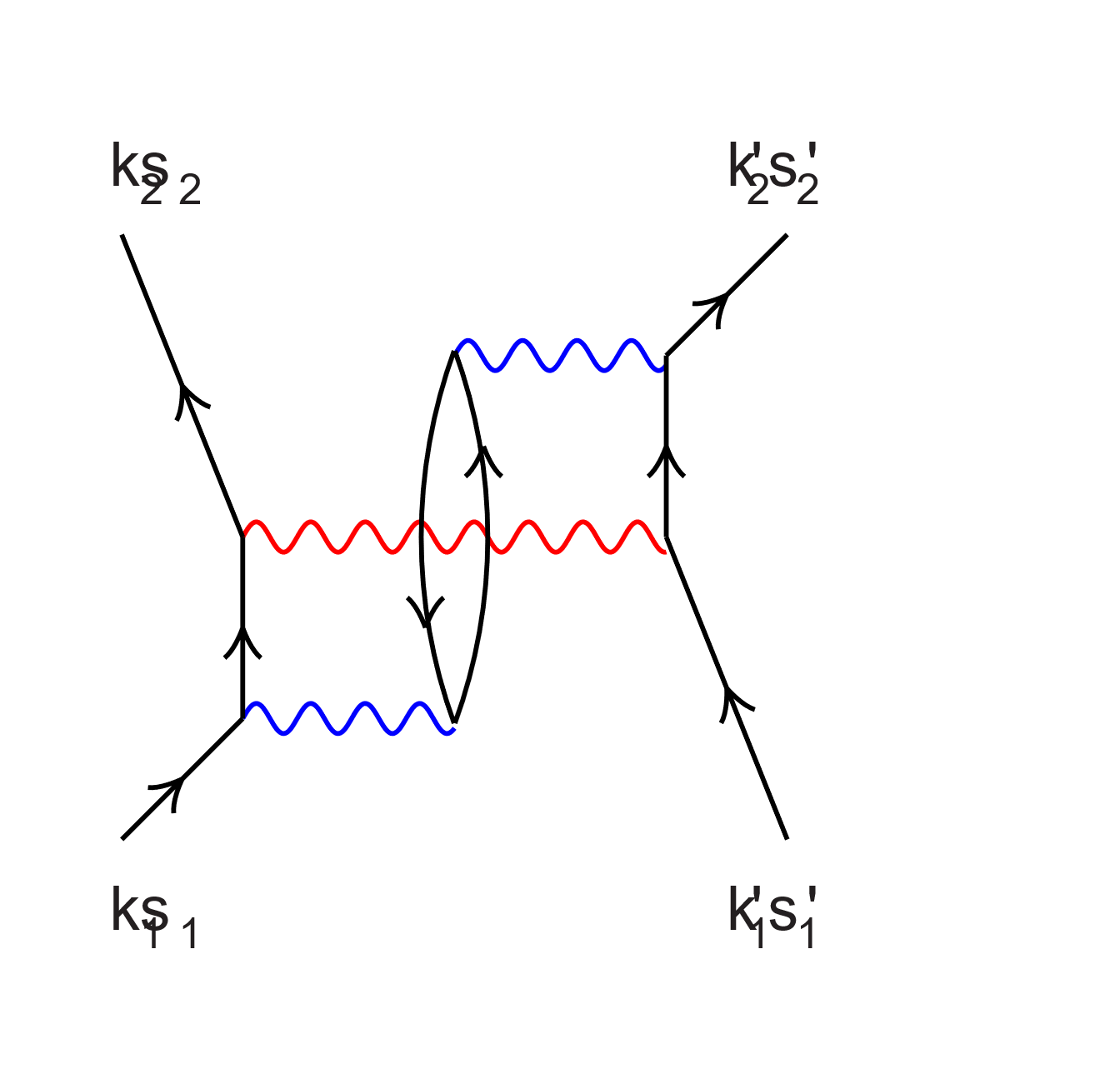}} 
\mbox{~}\\[-3mm] 
\mbox{~}\hspace{8mm} (c) \\[-10mm]  \mbox{~}\hspace{8mm}  \includegraphics[width=6cm]{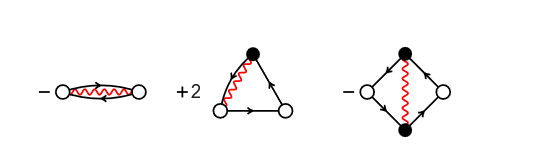}.
\caption{Examples of diagrams discussed in the text: (a) Parquet and (b) beyond-parquet diagram; (c) exchange diagrams.   
\label{fig_diagrams}
} 
\end{figure}

In correlated basis functions methods, the first step is to select a
trial $N-$body state $|\Phi_0\rangle$ that describes the system’s
statistics.  In the case of Fermionic systems, that could be a Slater
determinant of plane waves or other basis functions appropriate for
the symmetries of the system.  Next, one acts on the trial function
with an operator $\hat{F}_N$ that encodes dynamical
correlations. Among others, it suppresses the wave function at short
interparticle distances thus encoding short-range repulsion.  The correlated
and normalized wave function is
\[
|\Psi \rangle = \frac{\hat{F}_N |\Phi_0\rangle}{\langle \Phi_0 | \hat{F}_N^{\dagger} \hat{F}_N | \Phi_0 \rangle^{1/2} }  . 
\]
If $|\Phi_0\rangle$ is chosen to be a BCS state, the correlated-basis
function method leads to a BCS-like theory, with
the bare interaction replaced by an effective interaction which
includes many-body correlations.  The effective interaction is then
calculated by a systematic summation of parquet diagrams of rings and
ladders such as the one depicted in Fig.~\ref{fig_diagrams}(a)
and including exchange diagrams in the case of Fermion systems such as
the ones depicted in Fig.~\ref{fig_diagrams}(c). The
correlation operators are determined by an unconstrained
variational principle of the energy expectation value.

This well-known formalism encounters challenges in the case of the
nuclear interaction which has a rich operator structure with different
radial dependence for each term. Specifically, the bare nuclear
interaction is strongly spin-dependent with the spin channels
differing significantly. The necessary symmetrization of non-commuting
operators in the correlation operator $\hat F_N$ leads to terms which
we can identify with non-parquet diagrams. These contributions beyond
the parquet class describe processes where a spin flip can happen in
the intermediate states such as the one depicted in
Fig.~\ref{fig_diagrams}(b).  A systematic way to sum the
beyond-parquet diagrams has been developed~\cite{KrW2020}.  Here, we
present results for triplet pairing in neutron matter.

\section{Results} 
\label{results}
\begin{figure}[b]
\centerline{
\includegraphics[width=7.5cm]{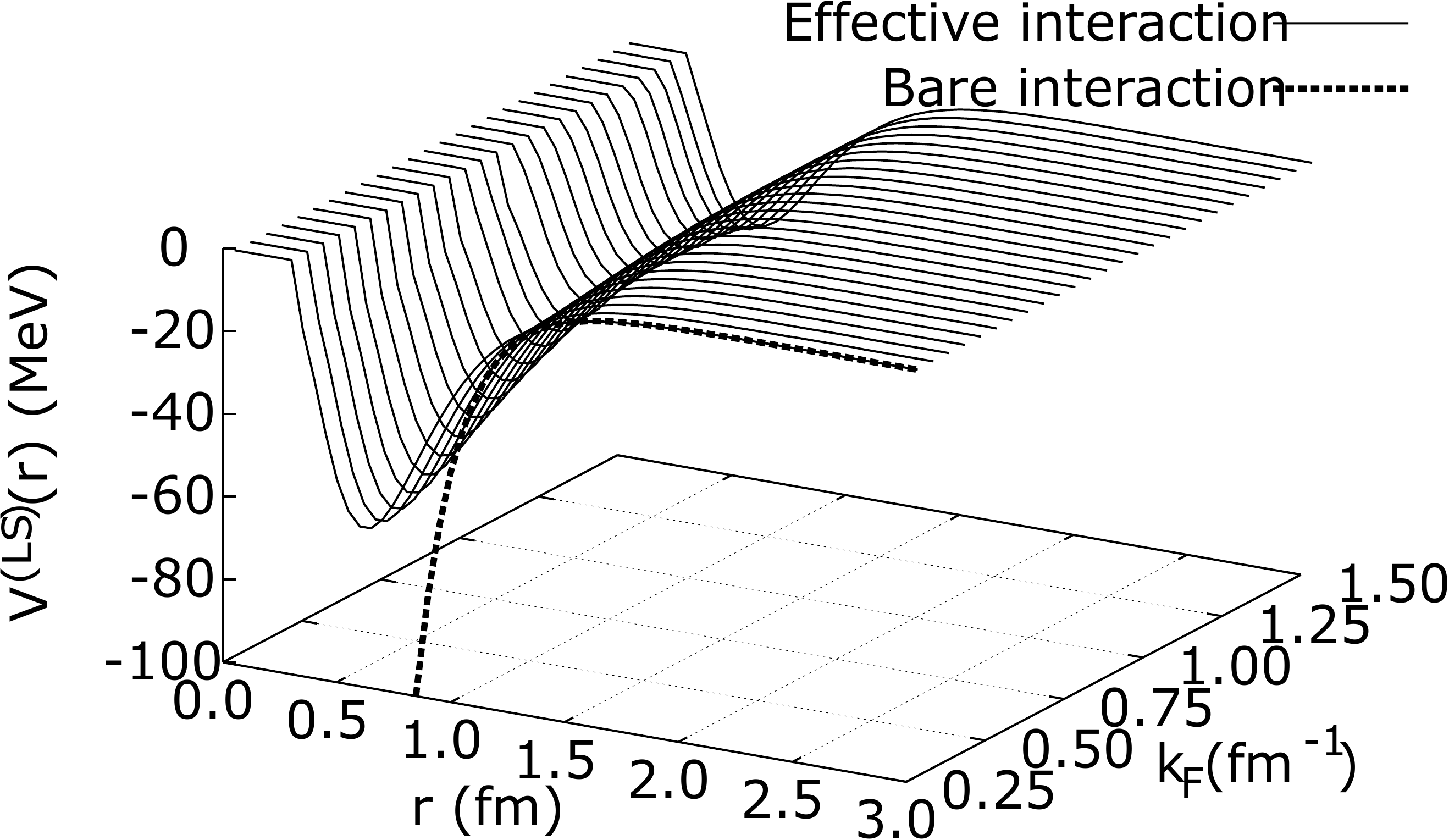} 
}
\caption{The in-medium spin-orbit effective interaction for the Reid $v_8$ interaction at different values of Fermi momentum (solid lines) compared with the bare interaction (dashed line). 
\label{fig_3d}
} 
\end{figure}

\begin{figure}[h]
\includegraphics[width=7cm]{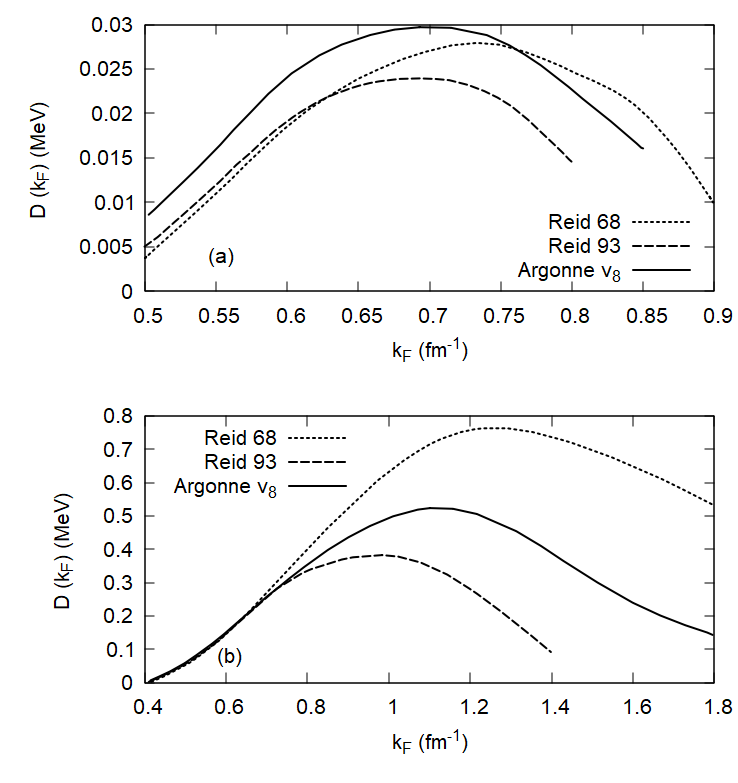}
\caption{(a) The $^3P_2- ^3\!F_2$ gap and (b) the $^3P_0$ gap calculated in this work with the indicated interactions. 
\label{fig_res}
} 
\end{figure}

In-medium effective interactions are calculated as outlined above for different nuclear potentials of $v_8$ type~\cite{KPW2023,KPW2024}.  
A dramatic suppression of the in-medium spin-orbit interaction is found, as shown in Fig.~\ref{fig_3d} in the case of the Reid $v_8$ potential. 
As a result, the $^3P_2- ^3\!F_2$ gap is found very weak, as shown in
Fig.~\ref{fig_res}(a).  In addition, the $^3P_0$ gap is enhanced, reaching
$0.4-0.8$~MeV for the interactions used here, as shown in
Fig.~\ref{fig_res}(b).

\section{Summary} 
\label{summary} 
We have carried out a comprehensive evaluation of the pairing
interactions in neutron matter, including medium polarization effects
and spin-flip processes, which so far have been ignored or treated in
simplistic phenomenological ways.  The most striking result is that
many-body correlations radically suppress the $^3P_2$ and
$^3P_2- ^3\!F_2$ gap by significantly reducing the in-medium spin-orbit
interaction. The $^3P_0$ pairing is found non-negligible.  Future
extensions could include the effect of three-body forces.
\noindent
\\
{\bf Acknowledgments} This work was supported by the Institute for Basic Science (2013M7A1A1075764).

%

\begin{thebibliography}{}
%
%
\bibitem{SeC2019}
A. Sedrakian and J. W. Clark, Superfluidity in nuclear systems and neutron stars. Eur. Phys. J. A \textbf{55}, 167 (2019). 
\url{https://doi.org/10.1140/epja/i2019-12863-6}
\bibitem{Str2018}
G. C. Strinati, The BCS–BEC crossover: From ultra-cold Fermi gases to nuclear systems. Phys. Rep. \textbf{738}, 1 (2018), 
\url{https://doi.org/10.1016/j.physrep.2018.02.004} 
\bibitem{KrW2020}
E. Krotscheck and J. Wang, Variational and parquet-diagram calculations for neutron matter. II. Twisted chain diagrams. Phys. Rev. C \textbf{102}, 064305 (2020). \url{https://doi.org/10.1103/PhysRevC.102.064305}
\bibitem{KPW2023}
E. Krotscheck, P. Papakonstantinou, J. Wang, Triplet Pairing in Neutron Matter. Astroph. J. \textbf{955}, 76 (2023). 
\url{https://doi.org/10.3847/1538-4357/acee7c} 
\bibitem{KPW2024}
E. Krotscheck, P. Papakonstantinou, J. Wang, Variational and parquet-diagram calculations for neutron matter. V. Triplet pairing. Phys. Rev. C \textbf{109}, 015803 (2024). 
\url{https://doi.org/10.1103/PhysRevC.109.015803} 
\end{thebibliography}
%
%

\end{document}